\documentclass{PoS}

\newcommand{\tmtexttt}[1]{{\ttfamily{#1}}}

\usepackage[square,sort,comma,numbers]{natbib}
\RequirePackage{nccmath}
\RequirePackage{wrapfig} % For figures next to text

\RequirePackage{graphicx}
\RequirePackage{subfig}
\RequirePackage{float}
\RequirePackage{amsmath}
\RequirePackage{verbatim}
\RequirePackage{rotating}
\RequirePackage{url}
\RequirePackage{acronym}
\RequirePackage{enumerate}
\RequirePackage{caption}
\RequirePackage{multirow}
\usepackage{url}
\setcitestyle{square}
\usepackage{color}

\title{Layout design studies for medium-sized telescopes within the Cherenkov Telescope Array}

\ShortTitle{Layout design studies for medium-sized telescopes within the Cherenkov Telescope Array}

\author{\speaker{T. Hassan}$^{1}$, B. Humensky$^{2}$, D. Nieto$^{2}$, M. Wood$^{3}$ for the CTA Consortium\footnote{Full consortium list at http://cta-observatory.org}\\
        $^{1}$IFAE, Edifici Cn., Campus UAB, E-08193 Bellaterra, Spain\\
        $^{2}$Columbia University, Department of Physics\\
        $^{3}$SLAC National Accelerator Laboratory, Stanford University\\
        E-mail: \email{thassan@ifae.es}}
\abstract{\color{black}The Cherenkov Telescope Array (CTA) is an international project for a next-generation ground-based gamma-ray observatory. CTA, conceived as an array of tens of imaging atmospheric Cherenkov telescopes, comprising small, medium and large-size telescopes, is aiming to improve on the sensitivity of current-generation experiments by an order of magnitude and provide energy coverage from 20 GeV to more than 300 TeV. In this study we explore how the medium-sized telescopes layout design and composition impacts the overall CTA performance by analyzing Monte Carlo simulations including Davies-Cotton and Schwarzschild-Couder medium-sized telescopes.}

\FullConference{The 34th International Cosmic Ray Conference,\\
		30 July- 6 August, 2015\\
		The Hague, The Netherlands}

\begin{document}
\color{black}
\section{Introduction}

%TODO AÑADIR CITA A LA PAGINA DE CTA Y AL CTA CONCEPT!

% QUITAR REQUIREMENTS!!!! TAMBIEN DE LOS LABELS!!!!

The Cherenkov Telescope Array (CTA) \footnote{\url{http://www.cta-observatory.org/}} \cite{CTA_concept} represents the next generation of Imaging Atmospheric Cherenkov Telescopes (IACTs). The project consists of two ground-based facilities, one in each hemisphere, that will provide full sky coverage in the Very High Energy (VHE) $\gamma$-ray sky from some tens of GeV up to a few hundreds of TeV with unprecedented capabilities, improving on the sensitivity of current IACT observatories by over an order of magnitude. To achieve such a wide observable energy range, each observatory will contain some tens of IACTs of different sizes: few Large Sized Telescopes (LSTs) optimized to detect the faint low energy showers, a moderate number of Medium Sized Telescopes (MSTs) increasing the effective area of the array and the number of telescopes simultaneously observing each event (multiplicity) for energies from 100 GeV up to 10 TeV, and a large number of Small Sized Telescopes (SSTs), spread out over some km$^2$ to catch the rare events at the highest energies of the electromagnetic spectrum.

The Southern Hemisphere site will take advantage of its privileged location to observe the inner Galactic Plane and Galactic Center with $\sim 100$ telescopes, while the Northern Hemisphere site will complement it with a reduced number of telescopes ($\sim 20$), with access to most of the extragalactic sources detected at these energies and other interesting sources such as the Crab, Geminga, M82 and Tycho. The MSTs dominate the performance in the core energy range of CTA from 100 GeV up to 5 TeV, where the point-source sensitivity is background-limited. These telescopes will populate the central part of the CTA layouts to improve the background rejection power by increasing the event multiplicity. Around 15 MSTs are expected to be built in the Northern Hemisphere, increasing that number up to $\sim 50$ in the Southern site with a spacing optimizing the best trade-off between event quantity and reconstruction quality. 

\begin{figure}[!h]
\subfloat[DC-MST]{%
\begin{minipage}[c][1\width]{0.5\textwidth}%
\centering
\includegraphics[clip,width=0.65\textwidth]{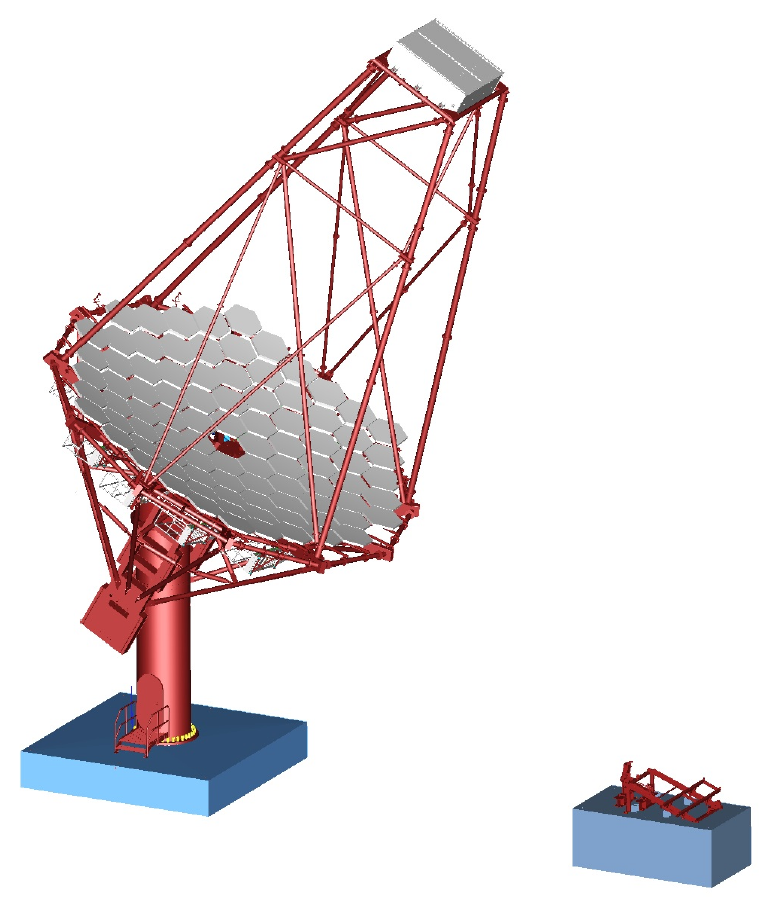}\label{fig:dcmst_baseline}%
\end{minipage}}\subfloat[SC-MST]{\centering{}%
\begin{minipage}[c][1\width]{0.5\textwidth}%
\begin{center}
\includegraphics[clip,width=0.55\textwidth]{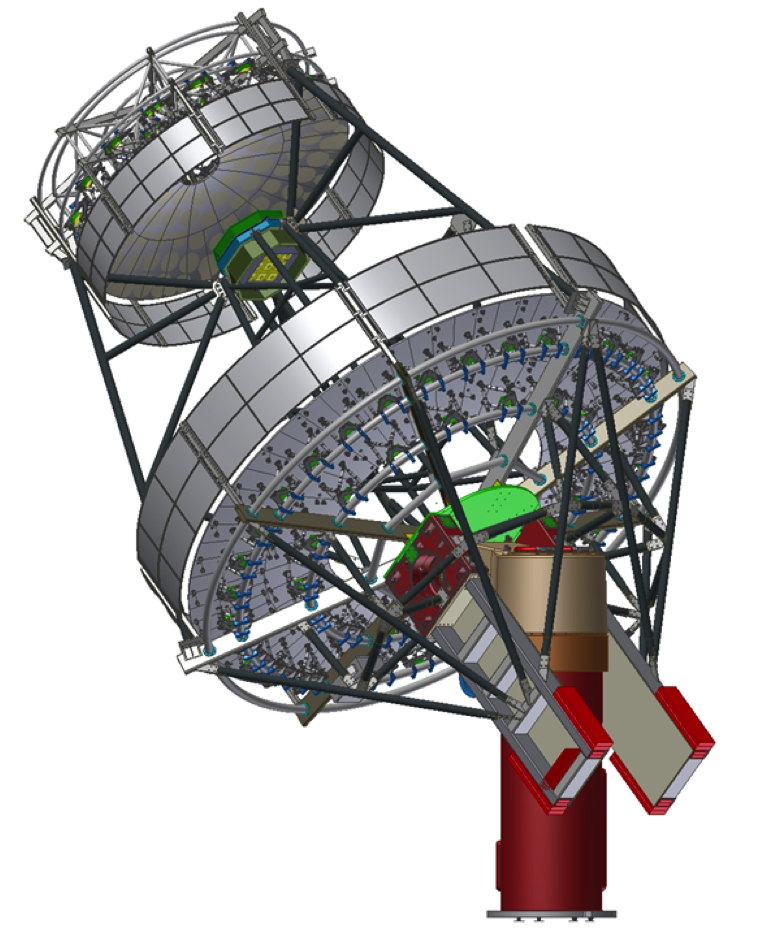}\label{fig:scmst_baseline}
\par\end{center}%
\end{minipage}}
\caption{Proposed MSTs by the CTA Consortium: \textit{Left}: Baseline design of the DC-MST, with 12 m diameter Davies-Cotton optics. \textit{Right}: Baseline design of the SC-MST, with a 9.7 m primary diameter Schwarzschild-Couder double-mirror optics.}
\end{figure}

Two designs for medium-sized telescope candidates have been proposed and developed:

\begin{itemize}

\item \textbf{The Davies-Cotton MST}: This telescope is a 12 m diameter single-mirror IACT (see Fig. \ref{fig:dcmst_baseline}\color{black}) built with a modified Davies-Cotton optics. This optical design improves its off-axis performance with respect to the traditional Davies-Cotton design by reducing the
optics-induced time spread to a negligible level. With a focal length of $16$ m it will have a field of view (FoV) of $\sim 8^{\circ}$ projected into a camera similar to the one of the LST, containing $\sim 1800$ Photo-Multiplier Tubes (PMTs).

\item \textbf{The Schwarzschild-Couder MST}: This telescope features a novel two-mirror optical design that fully corrects spherical and comatic aberrations while providing a large FoV. After reflecting from its 9.7 m diameter primary mirror the demagnification of the shower images performed by the secondary mirror provides a fine plate scale that allows for highly pixelized focal plane instrumentation. The current design includes a camera with 11328 Silicon photomultiplier (SiPM) pixels with a total FoV of $\sim 8^{\circ}$, capable of recording the shower development with an excellent image resolution.

\end{itemize}

%At the time this work was written, the Southern Hemisphere site was planned to be extended with a yet undefined number of SC-MSTs, the telescope designed by the North American groups belonging to the CTA Consortium. 
In this work, we compare the performance of the proposed MST designs, and study the most efficient approach of telescope allocation by analyzing data simulated within the second large-scale Monte Carlo (MC) production (Prod-2) \cite{MC_ICRC:2013, MC_ICRC:2015}. Note that none of the array layouts considered in this study are meant to represent the final CTA layout, but are rather utilized for comparison purposes.

\section{Monte Carlo simulations and analysis}
\label{sec:MCsim}

Similarly as described in \cite{MC_ICRC:2013,APP_CTA_MC}, Extended Air Showers (EASs) initiated by cosmic and $\gamma$-rays are simulated using CORSIKA \cite{corsika} in a wide area ($\sim$20 km$^2$) over 5 orders of magnitude in energy. A master array layout containing more than 200 telescopes is then simulated with the \tmtexttt{sim\_telarray} package \cite{Konrad:2008}. The analysis of different subsets of telescopes extracted from the master array allows a comparison of  the performance of layouts with diverse telescope configurations. Telescopes were selected from the CTA Prod-2, simulated for the Leoncito site (high altitude and moderate geomagnetic field intensity) and analyzed using the MAGIC Analysis and Reconstruction Software (MARS) \cite{MARS} (introduced in \cite{APP_CTA_MC}). 

Differential sensitivities and angular resolutions calculated in this contribution correspond to the averaged performance of two simulated positions in the sky, pointing into the North and South directions (0$^\circ$ and 180$^\circ$ in azimuth) at 20$^\circ$ in zenith angle. In addition, off-axis performance is evaluated using diffuse $\gamma$-rays (generated over a wide sky region rather than from a single point), by evaluating the sensitivity of a point-like source located at different distances from the center of the camera. We impose the following conditions during our analyses: five standard deviations (5 $\sigma$) at each energy bin, at least 10 excess events above background and a signal-to-background ratio larger than 0.05.

\begin{figure}[!h]
   \centering
    \includegraphics[width=0.8\textwidth]{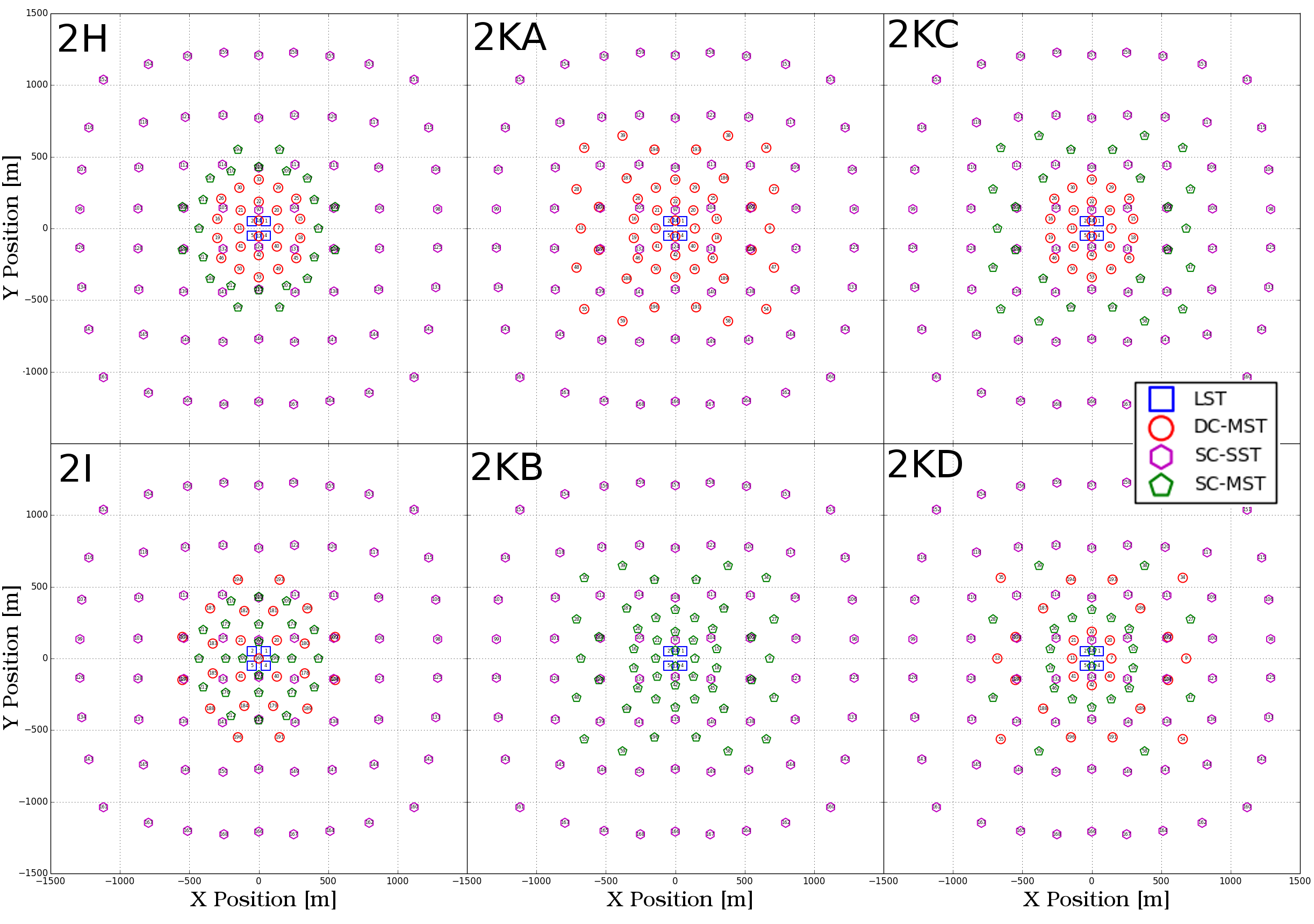}
    \caption[Proposed mixed DC/SC-MSTs layouts]{Proposed layouts to assess the impact of mixed MST types on CTA performance by applying different approaches to their location and spacing. Blue squares correspond to LSTs, red circles to DC-MSTs, pink hexagons to SC-SSTs and green pentagons to SC-MSTs. Telescope placement is tested using the \textit{halo} (``2H'' and ``2KC'') and the \textit{interleaved} (``2I'' and ``2KD'') approaches. Telescope spacing efficiency is also explored by comparing a \textit{compact} (``2H'' and ``2I'') with a \textit{graded} (``2I'' and ``2KD'') distribution.}
    \label{fig:prod2_SCT_layouts}
\end{figure}

We propose a set of 6 different layouts (shown in Fig. \ref{fig:prod2_SCT_layouts}\color{black}) to test the best approach to integrate the two different types of MSTs into the array. Several arrangement approaches can be tested with these layouts: the \textit{interleaved} placement (see ``2I'' and ``2KD'' subarrays), with both telescope types covering the central region of the array and the \textit{halo} option (``2H'' and ``2KC''), placing DC-MSTs in the center and SC-MSTs encircling them. The impact of the telescope spacing on the differential sensitivity is also tested by comparing a \textit{compact} distribution of telescopes (``2H'' and ``2I'') against a \textit{graded} one (``2KC'' and ``2KD''). In order to contrast DC-MST and SC-MST absolute performance, equivalent layouts with pure DC-MSTs (``2KA'') or SC-MSTs (``2KB'') are also proposed.

All considered layouts have equal number of LSTs (4) and 4m SC-SSTs (72) with identical positions and very similar number of DC-MSTs and SC-MSTs, ranging between 24 and 26 of each type (except for ``2KA'' and ``2KB'' layouts with 50 DC/SC-MSTs respectively). 

As a caveat, the current analysis methods may not be fully exploiting the potential of the fine resolution images captured by the highly pixelized SC-MSTs camera, as image cleaning algorithms and data quality cuts may not be optimal.

%As the SC-MST is the first designed IACT with such a densely pixelated camera, the current analysis methods may still not be tuned for such capabilities. In fact certain steps of the current MARS analysis, such as the standard 2-level image cleaning, are known to perform better for classic IACT cameras, such as the one built in the DC-MST. Alternative methods should be tested, such as the ``sum cleaning'' \cite{Rissi:2009} or ``aperture cleaning'' \cite{Wood:2014}, expected to be more efficient for the SC-MST.

\section{DC/SC-MST comparison studies}
\label{sec:Analysis}

\begin{figure}[!h]
  \centering
  \subfloat[Differential sensitivity]{\label{fig:diff_sens_SCTs_MSTs}\includegraphics[clip,width=0.40\textwidth]{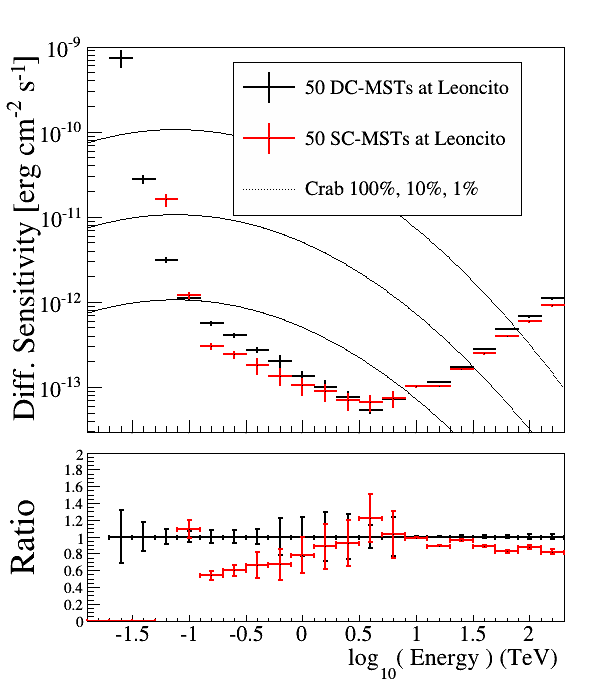}}
  \subfloat[Angular resolution]{\label{fig:ang_res_SCTs_MSTs}\includegraphics[clip,width=0.4\textwidth]{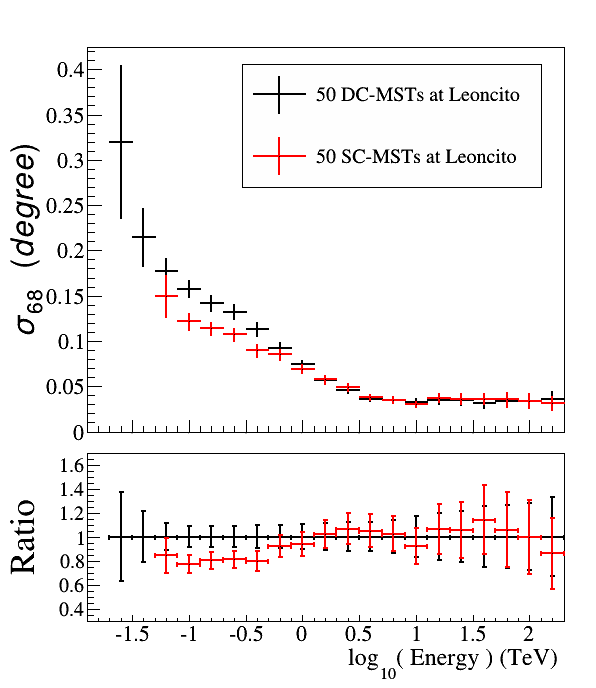}}     
  \caption[Differential sensitivity and angular resolution of DC-MSTs vs SC-MSTs.]{Comparison of the differential sensitivity in 50 hours (\textit{Left}) and the angular resolution (\textit{Right}) between 2 layouts of equal telescope distribution, one composed by 50 DC-MSTs and the other one by 50 SC-MSTs. Both layouts were simulated for the Leoncito site (north/south pointing average). DC-MSTs show better low energy capabilities while SC-MSTs outperform them above $\sim$ 120 GeV, mainly due to their improved angular resolution.}
  \label{fig:SCTs_MSTs}
\end{figure}

In order to explore the performance of single-type MST layouts, 2 sub-layouts were analyzed and compared: the first layout corresponds to the 50 DC-MSTs contained within layout ``2KA'' and the second one is the 50 SC-MSTs present in the candidate array ``2KB''. Both arrays have equal number of telescopes located at identical positions, each with a different MST telescope type. Figures \ref{fig:diff_sens_SCTs_MSTs}\color{black}~and \ref{fig:ang_res_SCTs_MSTs}\color{black}~show the differential sensitivity and angular resolution of these layouts simulated for the Leoncito site.

Although SC-MSTs do not help much below $\sim$ 100 GeV due to their smaller light collection area, they show good performance above that point, improving the sensitivity within the core energies of CTA up to a 50\% with respect to the DC-MSTs. A comparison of $\theta^2$ cut efficiencies and background rates attributes this improvement to the enhanced angular resolution of the SC-MSTs (due to the higher pixelization of the shower image), allowing a greater background rejection. Taking into account these results rely on an analysis that is not optimized for high resolution images (mainly tuned for classical IACTs) results are consistent with previous studies \cite{Tobias}. Further improvements of the analysis methods would very likely bring an improved SC-MST performance.

The next step is to compare both MST designs performance under a more reasonable scenario: with full layouts composed of the mixture of LSTs, MSTs and SSTs. For this reason, ``2KA'' and ``2KB'' layouts were analyzed. The low-energy performance of the DC-MSTs provides the ``2KA'' layout with a 30\% more differential sensitivity at 80 GeV than the layout ``2KB'', while the later differential sensitivity is boosted with respect to the former by the SC-MSTs up to a 25\% at 150 GeV.

%Results show the loss in the low energy performance caused by the smaller reflecting area of SC-MSTs (losses in the differential sensitivity of a 30\% at 80 GeV) is compensated with their improved reconstruction in the core energies (boost of a 25\% at 150 GeV). 

\section{MST spacing and distribution studies}
\label{subsec:sct_spacing}

Comparing the expected performance of the different proposed mixed MST type layouts, shown in Fig. \ref{fig:prod2_SCT_layouts}\color{black}, will guide us on the most efficient approach for MSTs arrangement. The optimum separation of MSTs is still to be decided. Larger inter-telescope distances would increase the effective area improving the chances of detecting the rare high-energy events, while closer telescopes would improve the low energy coverage, by increasing the multiplicity of IACTs observing each event. 

% TOO LONG!!!
%
%\begin{figure}[!h]
%\hfil
%\includegraphics[width=0.81\textwidth]{All_mixed_layouts_and_ratios.png}
%\hfil
%\caption[]{Differential sensitivity of mixed DC/SC-MST layouts (shown in Fig. \ref{fig:prod2_SCT_layouts}) in 50 hours of observation (average of north and south pointing directions) simulated at the Prod-2 Leoncito site. First two ratio plots compare layouts following the halo approach (``2H'' or ``2KC'') with the interleaved option (``2I'' or ``2KD''), showing halo approach improves the core energy sensitivity by nearly a factor 2. Comparing compact (``2H'' or ``2I'') against \textit{graded} arrays (``2KC'' or ``2KD'') show no clear preferred option.}
%\label{fig:mixed_layouts_diff_sens_ratios}
%\end{figure}

\begin{figure}[!h]
  \centering
  \subfloat[Differential sensitivity]{\label{fig:add bibliography bibtex latexmixed_layouts_diff_sens}\includegraphics[clip,width=0.5\textwidth]{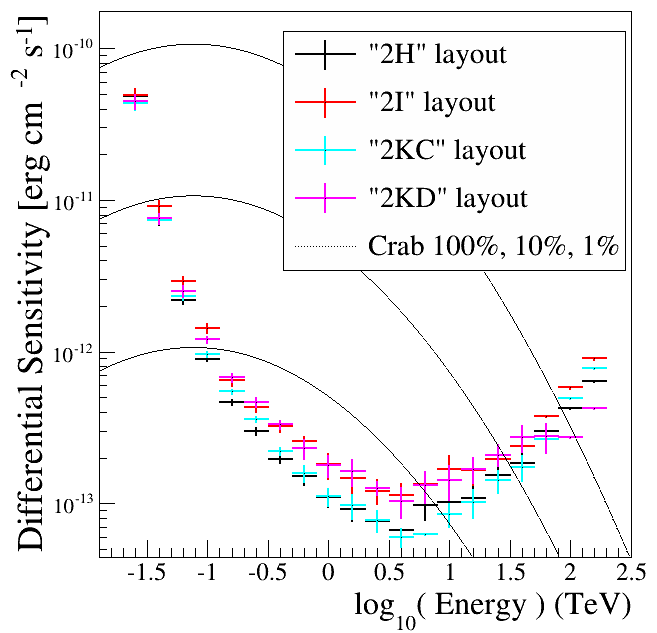}}
  \subfloat[Differential sensitivity ratios]{\label{fig:mixed_layouts_diff_sens_ratios_}\includegraphics[clip,width=0.5\textwidth]{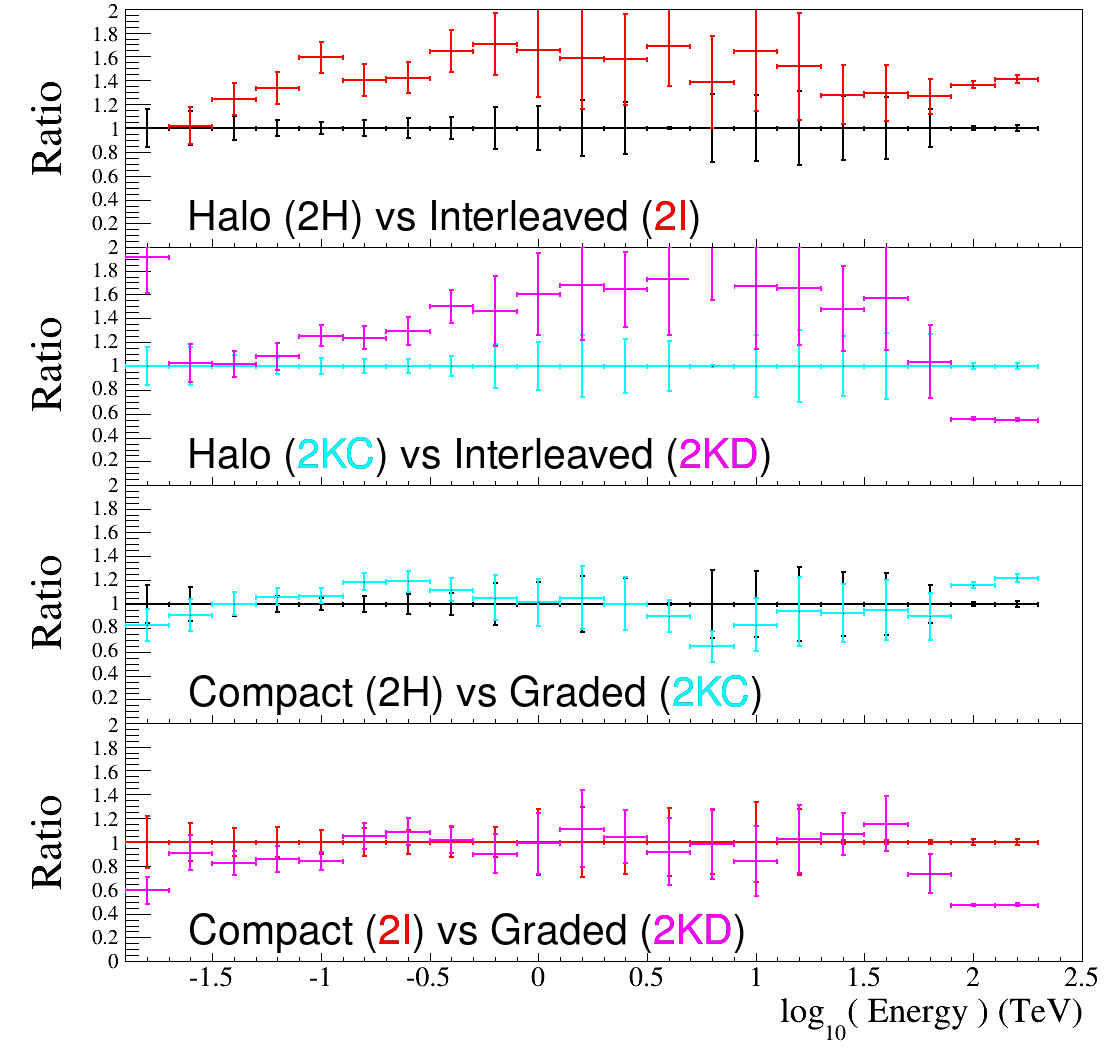}}     
  \caption[]{Differential sensitivity of mixed DC/SC-MST layouts (shown in Fig. \ref{fig:prod2_SCT_layouts}\color{black}) in 50 hours of observation simulated for the Prod-2 Leoncito site. The first two ratio plots compare layouts following the \textit{halo} approach (``2H'' or ``2KC'') with the \textit{interleaved} option (``2I'' or ``2KD''), showing that the \textit{halo} approach improves the core energy sensitivity by nearly a factor 2. Comparing \textit{compact} (``2H'' or ``2I'') against \textit{graded} arrays (``2KC'' or ``2KD'') does not show a clear preferred option.}
  \label{fig:mixed_layouts_diff_sens_ratios}
\end{figure}

\color{black}

Regarding the relative position of the different MST types in the layout, the \textit{halo} approach shows clear advantages over the \textit{interleaved} option. As shown in Fig. \ref{fig:mixed_layouts_diff_sens_ratios}\color{black}, comparison between ``2H'' over ``2I'' and ``2KC'' over ``2KD'' leaves no doubt about the sensitivity improvement of the \textit{halo} approach, up to a factor 2 in sensitivity. An interleaved configuration would imply that low energy event multiplicities will be reduced, as low-energy showers triggering the DC-MSTs may not necessarily trigger the SC-MSTs, thus negatively affecting their reconstruction. In addition, the amount of events reconstructed using both type of MSTs increases, weakening the effect of the better SC-MST image resolution, thus reducing the potential quality of the event direction reconstruction. The \textit{halo} approach maximizes the amount of low energy events well reconstructed between LSTs and DC-MSTs, and also increases the amount of higher energy showers reconstructed in stereo by the SC-MSTs.

The obtained results are not conclusive in regards to the determination of the most efficient telescope spacing. Comparing the \textit{compact} layouts (``2H'' and ``2I'') against the \textit{graded} ones (``2KC'' and ``2KD'') shows no clear preferred option. Observing the layouts using the preferred \textit{halo} approach, ``2H'' over ``2KC'' sensitivity ratio, a hint of improvement in the low energy range is seen by placing these telescopes closer to the center of the array, although the effect is balanced with a decreasing performance at higher energies. 

% No Cabe!!! :(

%\begin{figure}[!h]
%\hfil
%\includegraphics[width=0.81\textwidth]{Ang_res_all_mixed_layouts_and_ratios.png}
%\hfil
%\caption{Angular resolution (68\% containment) of mixed DC/SC-MST layouts (shown in Fig. \ref{fig:prod2_SCT_layouts}) in 50 hours of observation (average of north and south pointing directions) simulated at the Prod-2 Leoncito site. First two ratio plots compare layouts following the halo approach (``2H'' or ``2KC'') with the interleaved option (''2I'' or ``2KD''), showing halo approach also improves angular resolution ($\approx$ 60\%). Comparing \textit{compact} (``2H'' or ``2I'') against \textit{graded} arrays (``2KC'' or ``2KD'') show no clear preferred option.}
%\label{fig:mixed_layouts_ang_res}
%\end{figure}

%\begin{figure}[!h]
%   \centering
%    \includegraphics[width=11cm]{DiffSens_comparison.png}
%    \caption{Differential sensitivity of the CTA-N and CTA-S candidate arrays (50 hours of observation, N/S pointing average) together with the current generation of $\gamma$-ray detectors: \textit{Fermi}-LAT point-like sensitivity attained during 5 years of operation and MAGIC II and VERITAS differential sensitivity in 50 hours of observation.}
%    \label{fig:diffsens_comparison}
%\end{figure}

\section{MST off-axis performance studies}
\label{subsec:off_axis_performance}

Besides the performance on a steady point-like $\gamma$-ray source imaged at the center of the camera, off-axis capabilities are also crucial for a significant amount of CTA key scientific projects. Performing sky surveys or the detection of diffuse emission (for instance, emitted by hadronic interactions or DM signatures) relies on having a wide Field of View (FoV) and an efficient event reconstruction for off-axis showers. To characterize the observatory performance for different off-axis angles, the differential sensitivity of a point-like source located at different distances from the center of the camera is calculated.

To compare DC-MSTs with SC-MSTs, the off-axis capabilities of ``2KA'' and ``2KB'' layouts were analyzed. In addition, the off-axis performance of a mixed MST type layout was also tested, utilizing the layout showing better sensitivity in the previous section (``2KC''). As a figure of merit we focus on the radius $r_{50\%}$ defining the FoV region within which the off-axis sensitivity is better than a 50\% the on-axis sensitivity.

%\begin{figure}[!h]
%  \centering
%  \subfloat[2KA]{\label{fig:2KA_offaxis}\includegraphics[clip,width=0.4\textwidth]{Ratios_2KA.png}}
%  \subfloat[2KB]{\label{fig:2KB_offaxis}\includegraphics[clip,width=0.4\textwidth]{Ratios_2KB.png}}
%  \\
%  \subfloat[2KC]{\label{fig:2KC_offaxis}\includegraphics[clip,width=0.4\textwidth]{Ratios_2KC.png}} 
%  \caption[Off-axis sensitivity ratios of layouts with different MST types]{Relative off-axis sensitivity along different energy bins for 3 mixed MST types layouts (50h, N+S pointings at Leoncito site). \textit{Top left}: ``2KA'' layout, composed by 4 LSTs, 50 DC-MSTs and 72 SC-SSTs. \textit{Top Right}:``2KB'' layout, composed by 4 LSTs, 50 SC-MSTs and 72 SC-SSTs. \textit{Bottom}:``2KC'' layout, composed by 4 LSTs, 24 DC-MSTs, 26 SC-MSTs and 72 SC-SSTs.}
%  \label{fig:offaxis_rates_SCTs}
%\end{figure}

\begin{figure}[!h]
  \centering
  \subfloat[2KA]{\label{fig:2KA_offaxis}\includegraphics[clip,width=0.32\textwidth]{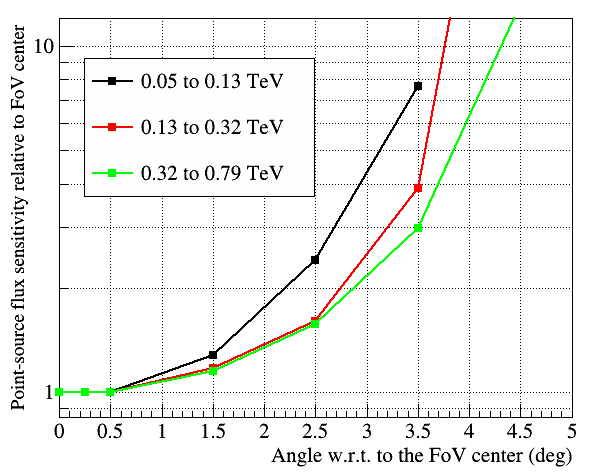}}
  \subfloat[2KB]{\label{fig:2KB_offaxis}\includegraphics[clip,width=0.32\textwidth]{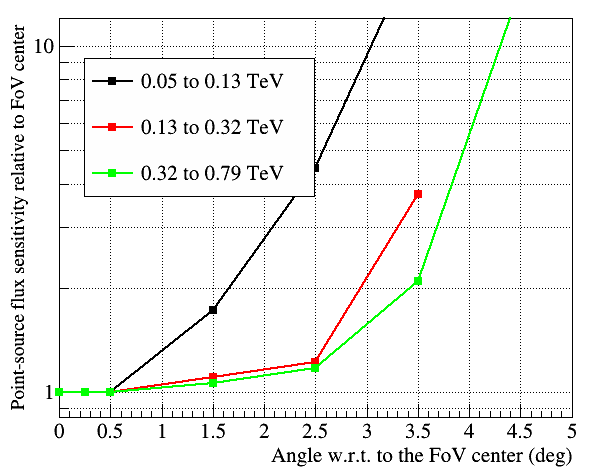}}
  \subfloat[2KC]{\label{fig:2KC_offaxis}\includegraphics[clip,width=0.32\textwidth]{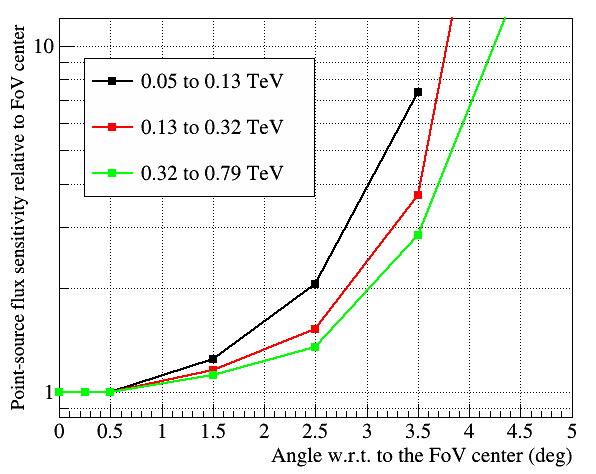}} 
  \caption[Off-axis sensitivity ratios of layouts with different MST types]{Relative off-axis sensitivity along different energy bins.}
  \label{fig:offaxis_rates_SCTs}
\end{figure}

Figure \ref{fig:offaxis_rates_SCTs}\color{black} ~shows off-axis sensitivity ratios for the three analyzed layouts. The following conclusions related to the off-axis performance of the different MST telescope types are inferred:

\begin{itemize}

\item The comparison between ``2KA'' and ``2KB'' shows that the low energy off-axis performance is enhanced by the DC-MSTs. Regarding energies higher that $\sim$130 GeV, SC-MSTs outperform the off-axis sensitivity contribution coming from the DC-MSTs, increasing $r_{50\%}$ by a 25\%.

\item The comparison between ``2KA'' and ``2KC'' shows that the central DC-MSTs in ``2KC'' are enough to attain a similar low energy off-axis performance as in ``2KA'' array, while the contribution of the \textit{graded} SC-MSTs improves $r_{50\%}$ by a 15\% at higher energies. 

\end{itemize}

\section{Conclusions}

% Esto es MUY políticamente incorrecto... Hablar mejor de que el halo approach es el mejor, y que los SCTs esperan mejorar la performance del core energies y la off-axis.

%Taking into account these conclusions, the most efficient layout for improving the core energy range of the CTA while augmenting its off-axis capabilities would be a mixed DC/SC-MST, following the \textit{halo} approach: DC-MSTs would be placed in the center near the LSTs while SC-MSTs would be placed surrounding them. Note this extension of SC-MSTs would greatly affect the amount of time needed for the different surveys the CTA Consortium is planning to perform, and would increase the chances of serendipitous detections through the observatory lifetime. 

The CTA project has triggered the design of new technologies to improve the IACTs capabilities. The Schwarzschild-Couder optics for the SC-MST show promising capabilities specifically in terms of angular resolution and off-axis performance. Concerning the telescope arrangement in an array potentially containing both types of MST, this study shows that the \textit{halo} approach outperforms the \textit{interleaved} option both in terms of sensitivity and angular resolution, providing specific guidelines on the future CTA baseline: for an array containing both types of MSTs the DC-MSTs should surround the core of LSTs while the SC-MSTs should encircle them, maximizing the amount of low-energy events reconstructed by LSTs and MSTs while improving the direction reconstruction and collection area of the showers with energies within the CTA core energy range. A mixed MST type array featuring a halo of 25 SC-MSTs surrounding a core of 25 DC-MSTs would boost the off-axis performance by a 15\% as compared to an array containing 50 DC-MSTs and no SC-MSTs, substantially improving CTA surveying capabilities and the chances of serendipitous discoveries.

%SC-MSTs would boost the off-axis performance by a 15\% as compared to an array containing only DC-MSTs as its medium-sized instruments, substantially improving CTA surveying capabilities and the chances of serendipitous discoveries.

\begin{acknowledgments}
The authors acknowledge the support of the Spanish MINECO under project FPA2010-22056-C06-06. We gratefully acknowledge support from the agencies and organizations  listed under Funding Agencies at this website: http://www.cta-observatory.org/.
\end{acknowledgments}

%\begin{thebibliography}{99}

%\end{thebibliography}
\bibliographystyle{JHEP}

\bibliography{skeleton} % References file

\providecommand{\href}[2]{#2}\begingroup\raggedright\begin{thebibliography}{1}

\bibitem{CTA_concept}
M.~{Actis} and et~al., {\it {Design concepts for the Cherenkov Telescope Array
  CTA}},  {\em Experimental Astronomy} {\bf 32} (Dec., 2011) 193--316,
  [\href{http://arxiv.org/abs/1008.3703}{{\tt arXiv:1008.3703}}].

\bibitem{MC_ICRC:2013}
K.~{Bernl{\"o}hr} et~al., {\it {Progress in Monte Carlo design and optimization
  of the Cherenkov Telescope Array}},  {\em ArXiv e-prints} (July, 2013)
  [\href{http://arxiv.org/abs/1307.2773}{{\tt arXiv:1307.2773}}].

\bibitem{MC_ICRC:2015}
T.~{Hassan} et~al., {\it {Second large scale Monte Carlo study for the
  Cherenkov Telescope Array}}, .

\bibitem{APP_CTA_MC}
K.~{Bernl{\"o}hr} et~al., {\it {Monte Carlo design studies for the Cherenkov
  Telescope Array}},  {\em Astroparticle Physics} {\bf 43} (Mar., 2013)
  171--188, [\href{http://arxiv.org/abs/1210.3503}{{\tt arXiv:1210.3503}}].

\bibitem{corsika}
D.~{Heck}, J.~{Knapp}, J.~N. {Capdevielle}, G.~{Schatz}, and T.~{Thouw}, {\em
  {CORSIKA: a Monte Carlo code to simulate extensive air showers.}}
\newblock {Forschungszentrum Karlsruhe GmbH}, Feb., 1998.

\bibitem{Konrad:2008}
K.~{Bernl{\"o}hr}, {\it {Simulation of imaging atmospheric Cherenkov telescopes
  with CORSIKA and sim\_ telarray}},  {\em Astroparticle Physics} {\bf 30}
  (Oct., 2008) 149--158, [\href{http://arxiv.org/abs/0808.2253}{{\tt
  arXiv:0808.2253}}].

\bibitem{MARS}
A.~{Moralejo} and {for the MAGIC collaboration}, {\it {MARS, the MAGIC Analysis
  and Reconstruction Software}},  {\em ArXiv e-prints} (July, 2009)
  [\href{http://arxiv.org/abs/0907.0943}{{\tt arXiv:0907.0943}}].

\bibitem{Tobias}
T.~{Jogler}, M.~D. {Wood}, J.~{Dumm}, et~al., {\it {Monte Carlo comparison of
  medium-size telescope designs for the Cherenkov Telescope Array}},  {\em
  ArXiv e-prints} (July, 2013) [\href{http://arxiv.org/abs/1307.5905}{{\tt
  arXiv:1307.5905}}].

\end{thebibliography}\endgroup

\end{document}